\documentstyle[12pt]{article}
\begin{document}
\begin{center}
{\Large \bf{Scale factor duality in string Bianchi cosmologies}}
\vspace*{10mm} \\
Elisa Di Pietro\footnote{E-mail: dipietro@astro.ulg.ac.be} and  
Jacques Demaret 
\vspace*{5mm} \\
Institute of Astrophysics and Geophysics \vspace*{3mm} \\
Group of Theoretical Cosmology \vspace*{3mm} \\
University of Li\`ege \vspace*{3mm} \\
B-4000 LIEGE-BELGIUM \vspace*{23mm} \\
\begin{abstract}
\noindent
We apply the scale factor duality transformations introduced in the context
of the effective string theory to the anisotropic Bianchi-type models. We 
find dual models for all the Bianchi-types [except for types $VIII$ and $IX$] 
and construct for each of them its explicit form starting from the exact 
original solution of the field equations. It is emphasized that the dual 
Bianchi class $B$ models require the loss of the initial homogeneity symmetry
of the dilatonic scalar field.
\end{abstract}
\end{center}
\hspace*{5mm} \vspace*{3mm} 
\newpage
\section{Introduction}
\hspace*{5mm}
String theory has recently motivated the study of cosmological models 
because its application to cosmology provides an alternative solution 
to the inflationary paradigm, the {\it pre-Big Bang scenario} \cite{prebb}.
Indeed, the low-energy string action possesses a symmetry property, called 
{\it scale factor duality}, which lets us expect that the present phase of 
the Universe evolution is preceded in time by a "naturally" inflationary 
pre-Big Bang phase. In such a scenario, the {\it Big Bang} represents only 
the peculiar instant of the Universe evolution in which its curvature and
its density are maximal. Unfortunately, as far as we know, it is not 
already possible in this context to avoid this maximum to be infinite and 
so, to remove the initial singularity present in the standard model.   

Several reasons lead us to think that the standard model characterized by
an isotropic and spatially homogeneous spacetime cannot be extended until
the first phases of the Universe \cite{bianchi}. Indeed, most cosmologists 
think that the primordial Universe was not necessarily isotropic: for them,
it seems more natural and more general to say that the Universe began in a 
less symmetric state and became isotropic after some time. So, more general 
spacetimes than FLRW models are often invoked to describe the real dynamical
behavior of the very early Universe. The cosmological spatially homogeneous 
and anisotropic spacetimes are the most symmetric models after those of FLRW
type. They are called the {\it Bianchi}-type models.
  
Since the pre-Big Bang scenario concerns principally the early Universe that 
has to be described by an anisotropic spacetime and since the scale factor 
duality is at the root of this scenario, it is important to demonstrate 
that it remains also valid for any anisotropic spacetime. The aim of this 
paper is to inquire about the possibility of building explicit exact dual 
solutions -- in the sense of scale factor duality -- for different Bianchi 
models in order to check the validity of the pre-Big Bang scenario in this 
more general cosmological framework.

The problem of the scale factor duality has already been considered within 
the context of spatially homogeneous vacuum Brans-Dicke cosmologies by 
Clancy {\it et al.} \cite{lidsey}. This study is based on the Lagrangian 
formulation of the field equations and is restricted to Bianchi type $A$ 
cosmological models, since it is well known that such formulation for class
$B$ models is ambiguous \cite{classb}. Moreover, the dilaton is supposed 
constant on the surfaces of homogeneity and the dual and the original 
equations are assumed to have the same isometry group, so that, in 
\cite{lidsey}, duality symmetries are found to exist for type $I$, $II$, 
$VI_{-1}$ and $VII_0$.

In the present paper, we adress this problem in the framework of effective 
string theory\footnote{Effective string theory with $H=0$ corresponds to a 
special case of Brans-Dicke cosmologies with $\omega=-1$, where $\omega$ is 
the Brans-Dicke parameter.}, in a more direct way. We rely on the 
construction
of the dual metric from the original one, using the explicit scale factor 
duality transformations as deduced from the study of O(d,d)-invariance of 
the effective action of string theory (\cite{prebb}, \cite{nous}). In 
particular, we give explicit expressions of the dual counterparts of the 
exact solutions known for different Bianchi models in presence of a 
dilatonic field \cite{batakis}.

With the exception of Bianchi class $A$ types $VIII$ and $IX$ models, we 
are able to construct explicit dual solutions for each Bianchi spacetime. 
However, for class $B$ models ($III$, $IV$, $V$, $VI_{h \not= -1}$ and 
$VII_{h \not= 0}$), we have to get rid of the hypothesis of spatial 
homogeneity of the dilatonic field and permit its effective inhomogeneity 
character.

\section{The SFD symmetry of the effective action of string theory}
\hspace*{5mm} 
In the four dimensional spacetime context, the low energy string effective 
action (in the string frame) can be written as\footnote{Greek indices always run 
from $0$ to $3$.} 
\begin{equation}
S_{eff} = \frac{1}{2 \kappa^2}\,\int{ e^{-\phi}\,\left[ R + 
\nabla_\alpha \phi \,\nabla^\alpha \phi 
- \frac{1}{12}\,H_{\alpha\beta\gamma}\,H^{\alpha\beta\gamma} \right] 
\sqrt{- g}\,\,d^4 x}
\label{action}
\end{equation}
Our application of the scale factor duality (SFD) to the Bianchi-type models 
will be made assuming
a vanishing antisymmetric field strength $H_{\alpha\beta\gamma}$. Within 
this assumption and by varying action (\ref {action}) with respect to the 
metric $g_{\alpha\beta}$ and to the dilatonic scalar field $\phi$, we find 
respectively the following field equations \cite{copeland}: 
\begin{equation}
R^{\alpha}_{\hspace{2mm}\beta} + g^{\alpha\delta} 
\nabla_\delta\nabla_\beta\,\phi = 0 
\label{champ1}
\end{equation}
and
\begin{equation}
R + 2\,\Box\,\phi - \left[ \nabla \phi \right]^2 = 0 
\label{champ2}
\end{equation}
where $\Box$ stands for the dalembertian operator.

It is well known that action (\ref {action}) is invariant under a SFD
transformation
\cite{prebb}. We have showed recently \cite{nous} that SFD has the same 
form for any kind of metric, i.e.
\begin{equation}
G \rightarrow \bar{G} = G^{-1} \hspace{14mm}
\label{dualg}
\end{equation}
\begin{equation}
\phi \rightarrow \bar{\phi} = \phi - ln(det\,G)
\label{dualphi}
\end{equation}
The only difference between a FLRW metric, an anisotropic metric or an
inhomogeneous metric appears in the building of the matrix $G$: it always
contains the metric components relative to the coordinates the metric does 
not depend on, the other components remaining unchanged after duality
transformations. We can express the only condition necessary for the
application of these transformations as follows: "If the metric depends on a 
particular $x^\alpha$ coordinate, then we must have 
\begin{equation}
g_{\alpha\beta} = 0
\label{condition}
\end{equation}
\noindent 
for all $\beta \not = \alpha$" \cite{nous}. For example, if the metric 
depends explicitly on $x^0$ and $x^1$ and does not depend on $x^2$ and 
$x^3$, then we can use the transformations (\ref {dualg}) and (\ref 
{dualphi}) only if the spacetime metric can take the following form:
$$
g_{\alpha\beta} = \left(
\begin{tabular}{cc|c}
$g_{00}$ & $0$ & $0$ \\
$0$ & $g_{11}$ & $0$ \\
\hline
$0$ & $0$ & $G$
\end{tabular}
\right)
$$
where $G$ is the following $2 \times 2$ matrix:
$$
G = \left(
\begin{tabular}{cc}
$g_{22}$ & $g_{23}$ \\
$g_{32}$ & $g_{33}$
\end{tabular}
\right)
$$

\section{SFD in Bianchi-type models}
\hspace*{5mm}
Bianchi-type models are anisotropic and spatially-homogeneous models in 
which a three-dimensional Lie group of isometries acts simply-transitively 
on the hypersurfaces of homogeneity (for an introduction to the anisotropic 
cosmologies, see e.g. \cite{maccallum}). They are nine in number but their
classification permits to split them into two classes\footnote{In what 
follows, we shall use the Bianchi models classification presented by Ryan and
Shepley in \cite{ryan}.}: there are six models in
the class~A ($I$, $II$, $VI_{-1}$, $VII_0$, $VIII$ and $IX$) and five in the 
class~B ($III$, $IV$, $V$, $VI_h$ and $VII_h$)\footnote{The 
Bianchi-type model called $VI_h$ (resp. $VII_h$) corresponds to a class B
model only for $h \not= -1$ (resp. $h \not= 0$): for $h=-1$ (resp. $h=0$), 
it becomes a class $A$ model.}. As we shall see, for some of these models, 
the dual dilatonic field loses the spatial homogeneity symmetry of the 
initial field. This is due to the fact that, in some cases\footnote{In 
fact, in all the class $B$ models.}, the determinant of the matrix $G$ 
becomes dependent on a spacelike coordinate and so, in view of (\ref 
{dualphi}), the initial homogeneous dilaton is transformed in an 
inhomogeneous field after SFD. As will be shown later, this loss of 
symmetry does not affect the spacetime metric which still keeps its initial
symmetry after duality.

The nine Bianchi-type metrics can all be written as follows if we adopt the
hypothesis of diagonality of the spatial metric
\begin{equation}
ds^2 = - \left( \tilde{\omega}^0 \right)^2 
+ a^2 \left( \tilde{\omega}^1\right)^2
+ b^2 \left( \tilde{\omega}^2\right)^2
+ c^2 \left( \tilde{\omega}^3\right)^2
\end{equation}
where $a$, $b$ et $c$ are the scale factors, functions of the timelike
coordinate only and where the $\tilde{\omega}^\alpha$ (with 
$\alpha=0,1,2,3$) are the Cartan 1-forms characterizing the different 
Bianchi metrics. But, as we want to apply relations introduced previously 
in \cite{nous}, we will have to write each metric in its natural basis and 
so to develop Cartan 1-forms in terms of natural 1-forms. 

Most of the string cosmological exact solutions given in the literature are 
not written in terms of the proper time $t$ but rather in terms of the 
logarithmic time $\tau$ related to proper time by the following differential 
expression:
\begin{equation} \displaystyle
dt(\tau) = a(\tau)\,b(\tau)\,c(\tau)\,e^{- \phi(\tau)}\,d\tau  
\end{equation}
So, unless explicit mention, we shall write the metric and perform the 
duality transformations in terms of the logarithmic time and use the 
following coordinate system:  for any metric, we shall take $(x^0, x^1, x^2, 
x^3) = (\tau, x, y, z)$. Every initial solution will be noted $a$, $b$, $c$ 
and $\phi$ whereas the barred variables, $\bar{a}$, $\bar{b}$, $\bar{c}$ and 
$\bar{\phi}$, will always refer to a dual solution.

For some of the class $A$ spacetimes, similar results have been found by
Clancy {\it et al.} \cite{lidsey} in the framework of scalar-tensor
theories. In this paper,
the authors impose to the dilatonic scalar field to remain homogeneous after
duality and so they are not able to find a dual solution for each 
Bianchi-type model. Our method to find dual solutions appears however 
simpler and more direct than theirs.

We shall present in the next section a table with explicit exact solutions 
and their corresponding dual solutions for Bianchi classes A and B models.

\subsection{Bianchi $I$ model}
\hspace*{5mm}
The Bianchi $I$ metric describes the simplest spatially homogeneous
aniso\-tro\-pic model:
\begin{equation}
ds^2 = - (abc)^2 e^{- 2 \phi} d\tau^2 + a^2 dx^2 + b^2 dy^2 + c^2 dz^2
\end{equation}
As this metric depends on the timelike coordinate only, the matrix $G$
needed for the SFD transformation is the following $3 \times 3$ matrix:
\begin{equation}
G = \left(
\begin{tabular}{ccc}
$a^2$ & $0$ & $0$ \\
$0$ & $b^2$ & $0$ \\
$0$ & $0$ & $c^2$ 
\end{tabular}
\right)
\end{equation}
with determinant $(abc)^2$. From (\ref {dualg}), we see that the
dual metric can be obtained by simply inverting the matrix $G$:
\begin{equation}
\displaystyle
ds^2 = - (\bar{a}\bar{b}\bar{c})^2 e^{- 2 \bar{\phi}}\, d\tau^2 + 
\bar{a}^2 dx^2 + \bar{b}^2 dy^2 + \bar{c}^2 dz^2
\end{equation}
and using (\ref {dualphi}), the dual dilatonic scalar field can be written 
as 
\begin{equation}
\bar{\phi} = \phi - 2\,log(abc) = \phi + 2\,log(\bar{a}\bar{b}\bar{c})
\end{equation}
with, for the dual scale factors, 
\begin{equation}
\bar{a} = a^{-1}, \hspace*{3mm} \bar{b} = b^{-1} \hspace{3mm} and
\hspace{3mm} \bar{c} = c^{-1}
\end{equation}

\subsection{Bianchi $II$ model}
\hspace*{5mm}
As mentioned above, the 
 SFD transformations given by (\ref {dualg})
and (\ref {dualphi}) have to be applied to a metric written in its natural
basis. The Bianchi $II$ metric developed in terms of the natural basis 
1-forms can be written as:
\begin{equation}
\begin{tabular}{rl}
$ds^2$ & $= - (abc)^2 e^{- 2 \phi} d\tau^2 + a^2 (dy - x dz)^2 + b^2 dz^2 
+ c^2 dx^2$ \vspace*{1mm} \\
& $= - (abc)^2 e^{- 2 \phi} d\tau^2 + c^2 dx^2 + a^2 dy^2 - 2 x a^2 dy dz 
+ (a^2 x^2 + b^2) dz^2$
\end{tabular}
\end{equation}
A spacelike coordinate appears now explicitly in the metric despite
its spatial homogeneity, so it is necessary to take it into account in the
construction of the matrix $G$. As the metric depends on the two 
coordinates, $\tau$ and $x$, $G$ is the following $2 \times 2$ matrix:
\begin{equation}
G = \left(
\begin{tabular}{cc}
$a^2$ & $- x a^2$  \\
$- x a^2$ & $x^2 a^2 + b^2$  
\end{tabular}
\right)
\end{equation}
with $(ab)^2$ as determinant. We note that despite the presence
of the coordinate $x$ in $G$, its determinant remains time-dependent only. In
view of (\ref {dualphi}), the dual dilatonic field
remains spatially homogeneous as the dual metric.

In order to retrieve the initial Cartan 1-forms
\begin{center}
\begin{tabular}{rcl}
$\omega^0$ & $=$ & $a\,b\,c\,e^{-\phi}\,d\tau$ \\
$\omega^1$ & $=$ & $dy - x dz$ \\
$\omega^2$ & $=$ & $dz$ \\
$\omega^3$ & $=$ & $dx$ 
\end{tabular}
\end{center}
in the dual metric, we have 
to add to the inversion of the matrix $G$ the following variable change:
\begin{center}
\begin{tabular}{rcl} 
$x$ & $\rightarrow$ & $-x$ \\
$y$ & $\rightarrow$ & $z$ \\
$z$ & $\rightarrow$ & $y$
\end{tabular}
\end{center}
and so, the dual metric takes the following form:
\begin{equation}
ds^2 = - (\bar{a}\bar{b}\bar{c})^2 e^{- 2 \bar{\phi}}\, d\tau^2 + 
\bar{a}^2 (dy - x dz)^2 + \bar{b}^2 dz^2 + \bar{c}^2 dx^2
\end{equation}
with the dual scale factors defined by 
\begin{equation}
\bar{a} = b^{-1}, \hspace{3mm} \bar{b} = a^{-1} \hspace{3mm} and \hspace{3mm}
\bar{c} = c
\label{cond2}
\end{equation}
The relation (\ref {dualphi}) and the definitions (\ref {cond2}) enable one
to write the dual dilaton in terms of the dual scale factors as follows:
\begin{equation} 
\bar{\phi} = \phi - 2\,log(ab) = \phi + 2 \,log(\bar{a}\bar{b})
\end{equation}

\subsection{Bianchi $III$ model}
\hspace*{4mm}
The Bianchi $III$ metric is a class $B$ spacetime with the following metric:
\begin{equation}
ds^2 = - (abc)^2 e^{- 2 \phi} d\tau^2 + a^2 dx^2 + b^2 dy^2 
+ c^2\,e^{2 x}\, dz^2 
\end{equation}
As it depends explicitly on two coordinates, $\tau$ and $x$, the matrix $G$ is
again a $2 \times 2$ matrix:
\begin{equation}
G = \left(
\begin{tabular}{cc}
$b^2$ & $0$ \\
$0$ & $c^2\,e^{2 x}\,$  
\end{tabular}
\right)
\end{equation}
with determinant $(bc)^2\,e^{2 x}$. To retrieve the initial Cartan 1-forms, it
is again necessary to add the transformation given by (\ref {dualg}) and 
(\ref {dualphi}) the following transformation on $x$:
$x \rightarrow - x$, with the following result for the dual metric:
\begin{equation}
\displaystyle
ds^2 = - (a\,b\,c)^2 e^{- 2\,\phi}\, d\tau^2 + 
\bar{a}^2 dx^2 + \bar{b}^2 dy^2 + \bar{c}^2\,e^{2 x} dz^2
\end{equation}
with 
\begin{equation}
\bar{a} = a, \hspace*{3mm} \bar{b} = b^{-1} \hspace{3mm} and 
\hspace{3mm} \bar{c} = c^{-1}
\label{cond3}
\end{equation}
and for the dual dilaton
\begin{equation}
\bar{\phi} = \phi - 2\,log(bc) + 2 x
\label{phidual3}
\end{equation}
This is an example of a dilatonic field becoming inhomogeneous after SFD
transformations due to the dependence of the determinant of $G$ with respect
to $x$.

Using (\ref {cond3}) and (\ref {phidual3}), we can write
\begin{equation} \displaystyle
a\,b\,c\,e^{-\phi}= \bar{a}\,\bar{b}\,\bar{c}\,e^{-\bar{\phi}+2 x}
\end{equation}
The expressions of the dual metric and the dual dilaton
in terms of the dual scale factors are finally given by:
\begin{equation}
\displaystyle
ds^2 = - (\bar{a}\bar{b}\bar{c})^2 e^{- 2 (\bar{\phi}-\,2\,x)}\, d\tau^2 + 
\bar{a}^2 dx^2 + \bar{b}^2 dy^2 + \bar{c}^2\,e^{2 x} dz^2
\label{abovemetric}
\end{equation}
\begin{equation}
\bar{\phi} = \phi + 2\,log(\bar{b}\bar{c}) + 2 x
\end{equation}
The 
inhomogeneity of the metric (\ref {abovemetric}) is only apparent, since
from (\ref {phidual3}), we can see that the exponential term 
$e^{- 2 (\bar{\phi}-\,2\,x)}$
present in the $g_{00}$ metric component is only $\tau$-dependent, so
that the dual metric remains spatially homogeneous after SFD transformations.

\subsection{Bianchi $IV$ model}
\hspace*{5mm}
The Bianchi $IV$ metric belongs to class B models. As the corresponding exact
solution we are going to present later is known in terms of the proper
time and is non-diagonal in Cartan's basis (cf. the table in the next 
section), we shall use the following form for its spacetime metric
\cite{bch4}:
\begin{equation}
ds^2 = - \sigma^0\sigma^0 + a^2 \sigma^1 \sigma^1 + b^2\,\left\{
c^2 \sigma^2 \sigma^2 + d^2 (\sigma^2 \sigma^3 + \sigma^3 \sigma^2) 
+ \sigma^3 \sigma^3 \right\}
\label{metric4}
\end{equation}
with $a$, $b$, $c$ and $d$, functions of $t$ and where the 1-forms
$\sigma^i$ ($i=0,1,2,3$) are defined by
\begin{equation}
\begin{tabular}{ll}
$\sigma^0 = dt$ \hspace{5mm} & $\sigma^2 = e^{-x}\,dz$ \vspace*{2mm} \\
$\sigma^1 = dx$ & $\sigma^3 = e^{-x}\,(dy -x dz)$
\end{tabular}
\label{sigma}
\end{equation}
Developing this metric in terms of the natural basis 1-forms, we obtain:
 \begin{equation}
ds^2 = - dt^2 + a^2 dx^2 + b^2\,e^{-2x}\,\left\{
dy^2 + \left[c^2 - 2\,x\,d^2 + x^2 \right] dz^2 + 2\,(d^2 - x)\,dy dz 
\right\}
\end{equation}

Again we can see that the metric depends on two coordinates, $t$ and $x$,
so that we take for $G$ the following
$ 2 \times 2$ matrix: 
\begin{equation}
G = b^2\,e^{- 2 x}\, \left(
\begin{tabular}{cc}
$1$ & $d^2-x$  \\
$d^2-x$ & $c^2-2xd^2+x^2$  
\end{tabular}
\right)
\end{equation}
Its determinant depends again on the coordinate $x$:
$det\,G = b^4\,e^{- 4 x}\,(c^2-d^4)$. Inverting $G$ and making the variable 
change:
\begin{center}
\begin{tabular}{l}
$x \rightarrow - x$\\
$y \rightarrow z$ \\
$z \rightarrow y$
\end{tabular}
\end{center}
we build the following dual metric: 
\begin{equation} \displaystyle 
ds^2 = - \sigma^0\sigma^0 + \bar{a}^2 \sigma^1 \sigma^1 + \bar{b}^2\,\left\{
\bar{c}^2 \sigma^2 \sigma^2 - \bar{d}^2 (\sigma^2 \sigma^3 
+ \sigma^3 \sigma^2) + \sigma^3 \sigma^3 \right\}
\end{equation}
with 
\begin{equation}
\bar{a} = a, \hspace{3mm} \bar{b} = b^{-1}\,(c^2-d^4)^{-1/4}, \hspace{3mm} 
\bar{c} = c \hspace{3mm} and \hspace{3mm} \bar{d} = d
\label{cond4}
\end{equation}
and with the same 1-forms $\sigma^i$ ($i=0,1,2,3$) as defined in (\ref
{sigma}).  Using (\ref {dualphi}) and (\ref {cond4}), we can also write the 
dual dilatonic field as
\begin{equation} 
\bar{\phi} = \phi - 4 x - log\left[b^4\,(c^2-d^4) \right] 
= \phi - 4 x + 4 \,log(\bar{b})
\end{equation}

\subsection{Bianchi $V$ model}
\hspace*{5mm}
The Bianchi $V$ metric developed in terms of its natural basis 1-forms can be
written as:
\begin{equation}
\displaystyle
ds^2 = - (abc)^2 e^{- 2 \phi} d\tau^2 + a^2 dx^2 + b^2\,e^{2 x}\, dy^2 
+ c^2\,e^{2 x}\, dz^2 
\end{equation}
The corresponding matrix $G$ is seen to be the following two-dimensional
squared matrix:
\begin{equation}
G = e^{2 x}\,\left(
\begin{tabular}{cc}
$b^2$ & $0$  \\
$0$ & $c^2$  
\end{tabular}
\right)
\end{equation}
with determinant $e^{4 x}\,(bc)^2$. This determinant being
$x$-dependent, it will lead to the inhomogeneity
of the dual dilatonic field.

The building of the dual metric needs both the inversion of the matrix $G$ and
the change: $x \rightarrow -x$. We can then write the dual
metric and the dual dilaton respectively as follows:
\begin{equation}
ds^2 = - (\bar{a}\bar{b}\bar{c})^2 e^{- 2(\bar{\phi}-4x)}\, d\tau^2 + 
\bar{a}^2 dy^2 + \bar{b}^2 (dz - x dy)^2 + \bar{c}^2 dx^2
\end{equation}
\begin{equation} 
\bar{\phi} = \phi + 4 x + 2\,log(\bar{b}\bar{c})
\label{phidual5}
\end{equation}
with 
\begin{equation}
\bar{a} = a, \hspace{3mm} \bar{b} = b^{-1} \hspace{3mm} and \hspace{3mm}
\bar{c} = c^{-1}
\end{equation}
\noindent
Again the inhomogeneity of the metric above is only apparent.

\subsection{Bianchi $VI_h$ model}
\hspace{5mm} 
Developed in its natural basis, the Bianchi $VI_h$ metric can be written as
\begin{equation} \displaystyle
ds^2 = - (abc)^2 e^{- 2 \phi} d\tau^2 + a^2 dx^2 + b^2 \,e^{2\,h\,x}\,dy^2 
+ c^2\,e^{2\,x}\, dz^2
\end{equation}
For all $h \not= -1$, this metric is a Bianchi class $B$ spacetime. The 
particular case of $h = -1$ transforms this metric in a Bianchi class $A$
model. For the purpose of this paper, it is not necessary to consider those
cases separately. 

The metric being independent of $y$ and $z$, $G$ is
the following $2 \times 2$ matrix:
\begin{equation}
G = \left(
\begin{tabular}{cc}
$b^2\,e^{2\,h\,x}$ & $0$  \\
$0$ & $c^2\,e^{2\,x}$  
\end{tabular}
\right)
\end{equation}
with $e^{2\,x(h+1)}\,(bc)^2$ as determinant.

The dual metric can be obtained by performing both the inversion of $G$
and a transformation on $x$: $x \rightarrow - x$ and can thus be written as
\begin{equation} \displaystyle
ds^2 = - (\bar{a}\bar{b}\bar{c})^2 e^{- 2 \bar{\phi}}\,e^{4x(h+1)}\,d\tau^2 
+ \bar{a}^2 dx^2 + \bar{b}^2\,e^{2\,h\,x}\, dy^2 + \bar{c}^2\,e^{2x}\, dz^2
\label{dualmetric6}
\end{equation}
with 
\begin{equation}
\bar{a} = a, \hspace{3mm} \bar{b} = b^{-1} \hspace{3mm} and \hspace{3mm} 
\bar{c} = c^{-1}
\end{equation}
Using (\ref {dualphi}), the dual dilatonic scalar field takes the
following form
\begin{equation}
\bar{\phi} = \phi + 2\,x\,(h+1) + 2\,log(\bar{b}\bar{c})
\label{phidual6}
\end{equation}
Note that the dual
dilatonic field remains spatially homogeneous only for the special case 
$h = -1$, which corresponds to a class $A$ spacetime.

\subsection{Bianchi $VII_h$ model}
\hspace{5mm} 
We can write the Bianchi $VII_h$ metric in its natural basis as follows:
\begin{equation}
\begin{tabular}{lcl}
$ds^2$ & $=$ & $- (abc)^2 e^{- 2 \phi} d\tau^2  
+ a^2\,\left[ (X - k Y)\,dy - Y dz \right]^2 $ \vspace*{2mm} \\
& & $+ b^2\,\left[ Y dy + (X + k Y)\,dz \right]^2 
+ c^2 dx^2$ \vspace*{2mm} \\
& $=$ & $- (abc)^2 e^{- 2 \phi} d\tau^2 + c^2 dx^2 
+  \left[a^2\,(X-kY)^2 + b^2\,Y^2\right] dy^2
$ \vspace*{2mm} \\
& & $+ 2 \left[ b^2\,Y\,(X+kY) - a^2\,Y\,(X-kY) \right]\,dy dz$ 
\vspace*{2mm} \\
& & $+ \left[ a^2\,Y^2 + b^2(X+kY)^2 \right]\,dz^2$
\end{tabular}
\end{equation}
where
\begin{equation}
\begin{tabular}{rcl}
$X(x)$ & $=$ & $e^{-kx}\,cos(qx)$ \\
$Y(x)$ & $=$ & $\frac{1}{q}\,e^{-kx}\,sin(qx)$ \\
$q$ & $=$ & $\sqrt{1-k^2}$ \\
$k$ & $=$ & $h/2$
\end{tabular}
\label{definitions}
\end{equation}
\noindent
For all $h \not= 0$, this metric belongs to Bianchi class $B$  but for the
special case $h = 0$, it becomes a Bianchi class $A$ metric. Again the
purpose of this paper does not require to consider this particular case 
separately from the others.

The explicit presence of the coordinate $x$ in the metric implies $G$ to be
the following $2 \times 2$ matrix:
\begin{equation}
G = \left(
\begin{tabular}{cc}
$a^2\,(X-kY)^2+b^2\,Y^2$ & $b^2\,Y(X+kY)-a^2\,Y(X-kY)$  \\
$b^2\,Y(X+kY)-a^2\,Y(X-kY)$ & $b^2\,(X+kY)^2+a^2\,Y^2$  
\end{tabular}
\right)
\end{equation}
Using the definitions (\ref {definitions}), we can write the determinant of
$G$ as: $det(G) = a^2\,b^2\,e^{-4kx}$.

The dual metric can be obtained making both the inversion of $G$ 
and a transformation on the constant $k$: $k \rightarrow - k$, i.e.
$h \rightarrow -h$\hspace{1mm} \footnote{This transformation on $h$ is 
possible because the constant $h$, in our conventions \cite{ryan}, has to 
be comprised between $-2$ and $2$.}.
We can thus write
\begin{equation} 
\begin{tabular}{rl}
$ds^2 =$ & $- (\bar{a}\bar{b}\bar{c})^2 e^{- 2 \bar{\phi}}\,e^{- 8 k x} 
d\tau^2
+ \bar{a}^2\,\left[ (X - k Y)\,dy - Y dz \right]^2 $ \vspace*{2mm} \\
& $+ \bar{b}^2\,\left[ Y dy + (X + k Y)\,dz \right]^2 
+ \bar{c}^2 dx^2$
\end{tabular}
\label{dualmetric7}
\end{equation}
with 
\begin{equation}
\bar{a} = a^{-1}, \hspace{3mm} \bar{b} = b^{-1} \hspace{3mm} and \hspace{3mm} 
\bar{c} = c
\label{cond7}
\end{equation}
\noindent
Using the relation (\ref {dualphi}) and the definitions (\ref {cond7}), we 
can build the dual dilatonic field as follows:
\begin{equation}
\bar{\phi} = \phi - 4 k x - 2\,log(ab) = \phi - 4 k x + 
2\,log(\bar{a}\bar{b})
\label{dualphi7}
\end{equation}
\hspace*{5mm}
Again the inhomogeneity of the $g_{00}$ component in the metric (\ref 
{dualmetric7}) is only apparent. On the other hand, the dual dilaton remains 
spatially homogeneous only for $k=0$, i.e. for $h=0$ (class $A$ model). 
Indeed, for all $h~\not=~0$ (class $B$ model), we lose the initial spatial
homogeneity symmetry of the dilatonic field.

\subsection{Bianchi $VIII$ and Bianchi $IX$ models}
\hspace*{5mm}
These are both class $A$ models respectively given (in terms of the
natural 1-forms) by \cite{ryan}: 
\begin{itemize}
\item \underline{Bianchi $VIII$ metric:}
\begin{equation}
\begin{tabular}{rl}
$ds^2 =$ & $- (a b c)^2\,e^{- 2 \phi}\,d\tau^2 + a^2 \left[
2\,x\,dz + (1-2xz)\,dy \right]^2$ \vspace*{1mm} \\
& $+ b^2 \left[ dx + (x^2 - 1)\,dz + (x + z - z\,x^2)\,dy \right]^2$ 
\vspace*{1mm} \\
& $+ c^2 \left[ dx + (x^2 + 1)\,dz + (x - z - z\,x^2)\,dy \right]^2$
\end{tabular}
\end{equation}
\item \underline{Bianchi $IX$ metric:}
\begin{equation}
\begin{tabular}{rl}
$ds^2 = $ & $ - (a b c)^2\,e^{- 2 \phi}\, d\tau^2 + a^2 \left[
cos(x) dy + dz \right]^2$ \vspace*{1mm} \\
& $+ b^2 \left[- sin(z) dx + sin(x)\,cos(z) dy \right]^2$ \vspace*{1mm} \\ 
& $+ c^2 \left[ cos(z) dx + sin(x)\,sin(z) dy \right]^2$
\end{tabular}
\end{equation}
\end{itemize}
\noindent
\par
In these cases, we cannot apply SFD transformations as given by
(\ref {dualg}) and (\ref {dualphi}) in the building of the dual solutions 
because the conditions (\ref {condition}) are not realized. So 
the method presented in \cite{nous} does not enable one to build dual 
solutions for these two models.

Nevertheless, it is possible to determine a dual solution for these models
using Busher's relations given in \cite{busher} and which can be written, 
for $B = 0$, as follows:
\begin{equation}
\begin{tabular}{ll}
$\displaystyle\bar{g}_{yy} = \frac{1}{g_{yy}}$ & $
\displaystyle\bar{B}_{yy} = 0$ \\ \\
$\displaystyle \bar{g}_{\alpha\beta} = g_{\alpha\beta} - 
\frac{g_{y\alpha} g_{y\beta}}{g_{yy}}$ & 
$\displaystyle \bar{B}_{\alpha\beta} = 0$\\ \\
$\displaystyle \bar{g}_{y\alpha} = 0$ & 
$\displaystyle \bar{B}_{y\alpha} = \frac{g_{y\alpha}}{g_{yy}}$ \\
\end{tabular}
\end{equation}
where $y$ is, for the both models, the only coordinate the metric does not 
depend on and where $\alpha$ and $\beta$ stand for the coordinates 
the metric depends on, i.e. $\tau$, $x$ and $z$.

It is not necessary to write explicity the two dual solutions to see 
clearly that these SFD transformations introduce a torsion field $B$ which 
was absent initially but, above all, cancel the non-diagonal components of 
the metric so that we lose the initial symmetry of the metric after this SFD.
Thus, for these cases, the relevance of the SFD transformations is less 
evident than for the others.

\section{Exact dual solutions for Bianchi models}
\hspace*{5mm}
We shall present in the following table exact Bianchi-type solutions
of field equations (\ref {champ1}) and (\ref {champ2}) with
their dual expressions. In the first column, we shall note the Bianchi-type.
In the second and in the third columns, the initial exact
solution and its dual will be respectively displayed. All the explicit 
initial solutions presented in the second column come from Batakis and 
Kehagias paper's \cite{batakis} except for the Bianchi $IV$ solution 
which has been obtained by Harvey and Tsoubelis \cite{bch4}. After
any explicit solution, we also give the constraint on the constants present 
therein. Note that the quantities $N$, $p_i$ and $q_j$ (with i=1,2,3 and
j=1,2) appearing in the table are constants. Introducing both solutions and 
dual solutions in the field equations (\ref {champ1}) and (\ref {champ2}), we
have checked that they satisfy exactly these equations.

The Bianchi $VII_h$ as well as $VII_0$ general exact solutions of field 
equations (\ref {champ1}) and (\ref {champ2}) being not known, we have not 
been able to build their dual counterparts in explicit form. 

\newpage
\begin{center}
\begin{tabular}{|c|l|l|}
\hline 
Type & Solution & Dual solution \\
\hline \hline 
&& \\
$I$ 
& $a(\tau)^2 = \displaystyle e^{(p_1 + N)\tau}$ 
& $\bar{a}(\tau)^2 = \displaystyle e^{-(p_1 + N)\tau}$ \\
& $b(\tau)^2 = \displaystyle e^{(p_2 + N)\tau}$ 
& $\bar{b}(\tau)^2 = \displaystyle e^{-(p_2 + N)\tau}$ \\
& $c(\tau)^2 = \displaystyle e^{(p_3 + N)\tau}$ 
& $\bar{c}(\tau)^2 = \displaystyle e^{-(p_3 + N)\tau}$ \\
& $\phi(\tau) = N \tau$ 
& $\bar{\phi}(\tau) = - (2 N + p_1 + p_2 + p_3)\,\tau$  \\
&& \\
with & $\sum_{i<j} p_i\, p_j = N^2$ & \\
&& \\
\hline 
&& \\
$II$ 
& $a(\tau)^2 = \displaystyle X^{-1}\,e^{N \tau}$ 
& $\bar{a}(\tau)^2 = \displaystyle X^{-1}\,e^{-(2 p_1 + N)\tau}$ \\
& $b(\tau)^2 = \displaystyle X\,e^{(2 p_1 + N)\tau}$ 
& $\bar{b}(\tau)^2 = \displaystyle X\,e^{- N \tau}$ \\
& $c(\tau)^2 = \displaystyle X\,e^{(2 p_2 + N)\tau}$ 
& $\bar{c}(\tau)^2 = \displaystyle X\,e^{(2 p_2 + N)\tau} $ \\
& $\phi(\tau) = N \tau$ 
& $\bar{\phi}(\tau) = - (N + 2 p_1) \tau$ \\
&& \\
with & $X(\tau)=1/p_3\,cosh(p_3\,\tau)$ & $4 p_1 p_2 - p_3^2 = N^2$ \\
&& \\
\hline 
&& \\
$III$ 
& $a(\tau)^2 = \displaystyle p_1\,e^{(p_2 + N)\tau}\,sinh^{-2}(p_1\,\tau)$ 
& $\bar{a}(\tau)^2 = \displaystyle 
p_1\,e^{(p_2 + N)\tau}\,sinh^{-2}(p_1\,\tau)$  \\
& $b(\tau)^2 = \displaystyle p_1\,e^{(N - p_2)\tau}$ 
& $\bar{b}(\tau)^2 = \displaystyle 
p_1^{-1}\,e^{(p_2 - N)\tau}$ \\
& $c(\tau)^2 = \displaystyle 
p_1\,e^{(p_2 + N)\tau}\,sinh^{-2}(p_1\,\tau)$ 
& $\bar{c}(\tau)^2 = \displaystyle
p_1^{-1}\,sinh^2(p_1\,\tau)\,e^{-(p_2 + N)\tau}$  \\
& $\phi(\tau) = N \tau$ 
& $\displaystyle \bar{\phi}(\tau) = - N \tau + 2 x - 2\, log(p_1) $ \\
& & \hspace{13mm} $\displaystyle + 2\,log\left(sinh(p_1\,\tau)\right)$  \\
&& \\
with & $4\, p_1^2 - p_2^2 = N^2$ & \\
&& \\
\hline 
&& \\
$IV$ 
& $a(t)^2 = \displaystyle (a_0^{-1}\,t)^2$ 
& $\bar{a}(t)^2 = \displaystyle (a_0^{-1}\,t)^2$ \\
& $b(t)^2 = \displaystyle \left(a_0^{-1}\,t\right)^{2\,a_0}$ 
& $\bar{b}(t)^2 = \displaystyle \left(a_0^{-1}\,t\right)^{-2\,a_0}$ \\
& $c(t)^2 = \displaystyle
1+\left[ a_0 \,log\left(a_0^{-1}\,t\right)\right]^2$ 
& $\bar{c}(t)^2 = \displaystyle
1+\left[ a_0\,log\left(a_0^{-1}\,t\right)\right]^2$ \\
& $d(t)^2 = \displaystyle a_0\,log\left(a_0^{-1}\,t\right)$ 
& $\bar{d}(t)^2 = \displaystyle 
a_0\,log\left(a_0^{-1}\,t\right)$ \\
& $\phi(t) = 0$ 
& $\bar{\phi}(t) = \displaystyle
- 4 x - 4\,a_0\,log\left( a_0^{-1}\,t \right)$  \\
&& \\
with & $a_0 = 4/5$& \\
&& \\
\hline 
\end{tabular}
\end{center}
\newpage
\begin{tabular}{|c|l|l|}
\hline 
Type & Solution & Duale \\
\hline \hline
&& \\
$V$
& $a(\tau)^2 = \displaystyle \frac{q_1}{2}\,X\,e^{N \tau}$ 
& $\bar{a}(\tau)^2 = \displaystyle \frac{q_1}{2}\,X\,e^{N \tau}$ \\
&& \\
& $b(\tau)^2 = \displaystyle \frac{q_2}{2}\,X\,e^{(p_2 + N)\tau}$ 
& $\bar{b}(\tau)^2 = \displaystyle 
\frac{2}{q_2\,X}\,e^{-(p_2 + N)\tau}$  \\
&& \\
& $c(\tau)^2 = \displaystyle 
\frac{1}{2\,q_2}\,X\,e^{(- p_2 + N)\tau}$ 
& $\bar{c}(\tau)^2 = \displaystyle \frac{2\,q_2}{X}\,e^{(p_2 - N)\tau}$ \\
&& \\
& $\phi(\tau) = N \tau$ 
& $\bar{\phi}(\tau) = \displaystyle - 2\,log(X(\tau)) - N \tau
$ \\
&& \hspace*{13mm} $+ 4 x + 2\,log(2)$ \\
&& \\
with & $\displaystyle X(\tau) = p_1\,/\,sinh(p_1\,\tau)$ & 
$ 3\,p_1^2 - p_2^2 = N^2 $ \\
&& \\
\hline 
&& \\
$VI_{-1}$
& $a(\tau)^2 = q_1^2\,p_1\, 
exp\left[p_3^4 e^{2 p_2 \tau}\right] e^{(p_1 + N)\tau}$
& $\bar{a}(\tau)^2 = q_1^2\,p_1\,
exp\left[p_3^4 e^{2 p_2 \tau}\right] e^{(p_1 + N)\tau}$ \\
& $b(\tau)^2 = \displaystyle p_3^2\,p_2\, e^{(p_2 + N)\tau}$ 
& $\bar{b}(\tau)^2 = \displaystyle \left(p_3^2\,p_2\right)^{-1}\,
e^{-(p_2 + N)\tau}$ \\
& $c(\tau)^2 = \displaystyle p_3^2\,p_2\, e^{(p_2 + N)\tau}$ 
& $\bar{c}(\tau)^2 = \displaystyle \left(p_3^2\,p_2\right)^{-1}\,
e^{-(p_2 + N)\tau}$ \\
& $\phi(\tau) = N \tau$ 
& $\bar{\phi}(\tau) = - (N + 2 p_2) \tau - 2\, log\left(p_3^2\,p_2\right)$ \\
&& \\
with & $p_2^2 + 2 p_1\, p_2 = N^2$ & \\
&& \\
\hline 
&& \\
$VI_h$
& $a(\tau)^2 = \displaystyle q_1^2\, 
X^{(h^2+1)}\,e^{\left(\frac{h-1}{h+1}\,p_2 + N\right)\tau}$
& $\bar{a}(\tau)^2 = \displaystyle q_1^2\, 
X^{(h^2+1)}\,e^{\left(\frac{h-1}{h+1}\,p_2 + N\right)\tau}$ \\
& $b(\tau)^2 = \displaystyle q_2^2\,
X^{h(h+1)}\,e^{(p_2 + N)\tau}$ 
& $\bar{b}(\tau)^2 = \displaystyle q_2^{-2}\,
X^{-h(h+1)}\,e^{-(p_2 + N)\tau}$ \\
& $c(\tau)^2 = \displaystyle q_2^{-2}\,
X^{(h+1)}\,e^{(- p_2 + N)\tau}$ 
& $\bar{c}(\tau)^2 = \displaystyle q_2^2\,
X^{-(h+1)}\,e^{(p_2 - N)\tau}$ \\
& $\phi(\tau) = N \tau$ 
& $\bar{\phi}(\tau) = - N \tau + 2\,x\,(h+1)$ \\
&& \hspace*{15mm} $- (h+1)^2\,log[X(\tau)]$
\\
&& \\
with & $\displaystyle
X(\tau)=\left[\frac{h+1}{p_1}\,sinh(p_1\,\tau)\right]^{\frac{-2}{(h+1)^2}}$
& $\displaystyle
\frac{4\left(h^2+h+1\right)}{\left(h+1\right)^2}\,p_1^2 - p_2^2 = N^2 $ \\
&& \\
\hline
\end{tabular}
\newpage
\noindent
{\large \bf{Acknowledgments}}
\vspace*{2mm} \\ \noindent
This work was supported in part by Belgian Interuniversity Attraction Pole
P4/05 as well as by a grant from "Fonds National de la Recherche
Scientifique". 
\vspace*{2mm} \\
To the memory of my professor Jacques Demaret. It was a pleasure to 
work with him. He will be most deeply missed.

\end{document}